\documentclass[aps, prapplied, amsmath, amssymb, showkeys, longbibliography, reprint]{revtex4-1}

\usepackage{graphicx}% Include figure files
\usepackage{dcolumn}% Align table columns on decimal point
\usepackage{bm}% bold math

\begin{document}

\preprint{AIP/123-QED}

\title{Dynamics of a Ferromagnetic Particle Levitated Over a Superconductor}% Force line breaks with \\

\author{Tao Wang}
\email{taowang@berkeley.edu}
\affiliation{Department of Physics, University of California, Berkeley, California 94720-7300, USA}
\author{Sean Lourette}
\affiliation{Department of Physics, University of California, Berkeley, California 94720-7300, USA}
\author{Sean R. O'Kelley}
\affiliation{Department of Physics, University of California, Berkeley, California 94720-7300, USA}
\author{Metin Kayci}
\affiliation{DWI Leibniz Institute for Interactive Materials - RWTH AACHEN, Germany}
\affiliation{Department of Physics, University of California, Berkeley, California 94720-7300, USA}
\author{Y. B. Band}
\affiliation{Department of Chemistry, Department of Physics, Department of Electro-Optics, and the Ilse Katz Center for Nano-Science, Ben-Gurion University, Beer-Sheva 84105, Israel}
\author{Derek F. Jackson Kimball}
\affiliation{Department of Physics, California State University, East Bay, Hayward, California 94542-3084, USA}
\author{Alexander O. Sushkov}
\affiliation{Department of Physics, Boston University, Boston, Massachusetts 02215, USA}
\author{Dmitry Budker}
\affiliation{Helmholtz Institute Mainz, Johannes Gutenberg University, 55099 Mainz, Germany}
\affiliation{Department of Physics, University of California, Berkeley, California 94720-7300, USA}
\affiliation{Nuclear Science Division, Lawrence Berkeley National Laboratory, Berkeley, California 94720, USA}

\date{\today}% It is always \today, today,

\begin{abstract}
Under conditions where the angular momentum of a ferromagnetic particle is dominated by intrinsic spin, applied torque is predicted to cause gyroscopic precession of the particle. If the particle is sufficiently isolated from the environment, a measurement of spin precession can potentially yield sensitivity to torque beyond the standard quantum limit. Levitation of a micron-scale ferromagnetic particle above a superconductor is a possible method of near frictionless suspension enabling observation of ferromagnetic particle precession and ultrasensitive torque measurements. We experimentally investigate the dynamics of a micron-scale ferromagnetic particle levitated above a superconducting niobium surface. We find that the levitating particles are trapped in potential minima associated with residual magnetic flux pinned by the superconductor and, using an optical technique, characterize the quasiperiodic motion of the particles in these traps.
\end{abstract}
%\MSC[2010] 00-01\sep  99-00
\maketitle

\section{Introduction}

A single ferromagnetic particle floating in free space could be a valuable object for precision measurements of magnetic fields, torques, and rotations \cite{kimball2016precessing,prat2017ultrasensitive,band2018dynamics}. However, in order to experimentally realize such a system with maximum sensitivity in an Earth-bound laboratory, the ferromagnetic particle must be isolated from the environment while being supported against gravity via some form of near-frictionless suspension. Optical tweezers and electromagnetic traps are often used to levitate and manipulate particles \cite{ashkin1997optical,millen2015cavity}, but interactions with external fields may cause heating and/or perturbations that degrade sensitivity of such levitated ferromagnetic particles to torques and forces of interest. Another method of near-frictionless levitation is via the Meissner effect. Superconducting levitation has been widely studied, but usually with cm-scale magnets \cite{davis1988stability, hellman1988levitation}. Here we study superconducting levitation of a micron-scale magnet, which exhibits dynamic behavior distinct from that of cm-scale ferromagnets levitated above superconductors because of the much smaller levitation height that is comparable with the dimensions of flux-pinning sites \cite{treimer2012observation}.

Spin-exchange relaxation-free (SERF) atomic magnetometers and Superconducting Quantum Interference Device (SQUID) magnetometers have realized sub-femtotesla sensitivity for measurement times on the order of a second \cite{dang2010ultrahigh,storm2017ultra}; their fundamental sensitivities are given by the standard quantum limit (SQL) \cite{auzinsh2004can}. A precessing ferromagnetic needle magnetometer was proposed as a system that can, in principle, far surpass the SQL of a free ensemble of spins \cite{kimball2016precessing}. The sensitivity of the precessing ferromagnetic needle magnetometer is a result of rapid averaging of quantum noise through internal spin-lattice interactions combined with, in the absence of external torques, the conservation of the total angular momentum of the needle.

The main challenge for development of a high-sensitivity ferromagnetic needle magnetometer is the need to freely suspend the needle with minimal coupling to the environment. The Meissner effect is a promising method for near frictionless levitation of a ferromagnetic particle. However, flux pinning inside a superconductor can prevent a micron-scale magnet from being able to freely precess. Flux-pinning is particularly problematic when the size of the particle is comparable with the size of the flux pinning sites on the surface of the superconductor. Another issue is that flux pinning can affect the levitation height and even counteract the Meissner effect; micron-scale magnets can be attracted by the fields from the flux-pinning sites. Although in principle flux pinning is minimal for type-I superconductors, such as lead, they can be easily oxidized in the atmosphere, and the oxide can generate flux pinning. Type-II superconductors have a relatively large critical field $H_{c1}$ (depending on the temperature), and under conditions where the external magnetic field is below $H_{c1}$, type-II superconductors can exhibit behavior similar to that of type-I superconductors: the Meissner effect is dominant. 

In order for precession to be the dominant motion of a ferromagnetic needle, the total angular momentum of the needle must come primarily from the net intrinsic spin of the polarized electrons ($N \hbar \gg I \Omega$), where $N$ is the total number of the polarized electrons, $\hbar$ is the reduced Planck's constant, $I$ is the moment of inertia and $\Omega$ is the precession frequency \cite{kimball2016precessing}. If the angular momentum associated with the rotational motion of the needle lattice is much larger than the net intrinsic spin, the dominant motion of the needle is oscillation about the direction of the field as is commonly observed, for example, in a magnetic compass. If external torques on the needle (from, for example, a magnetic field) are too large, the needle can acquire a large rotational angular momentum that would make precession difficult to observe. For this reason, the external magnetic field needs to be shielded. For example, for a 10 $\mu$m length needle, the field needs to be smaller than 10 $\mu$G \cite{kimball2016precessing}. For a larger-size particle, the threshold magnetic field below which precession becomes the dominant motion will be even smaller. This is another advantage of using superconducting levitation as a method of near frictionless suspension: because of the Meissner effect, near the surface of the superconductor, external magnetic fields are strongly shielded.

When type-II superconductors and some type-I superconductors are cooled at non-zero magnetic field, flux pinning can be significant. In our experiment, the niobium was cooled with a magnetized ferromagnetic particle lying directly on the niobium surface. In future iterations of the experiment, we plan to implement techniques to load the ferromagnetic needle after the superconductor has been zero-field-cooled to reduce flux trapping. However, it is still difficult to eliminate all flux pinning in a superconductor even for a high-purity polycrystalline type-I superconductor \cite{treimer2012observation}. Therefore, studying and modeling the dynamics of a particle levitated over a superconductor in the presence of flux pinning is essential for understanding the dynamic behavior of the levitated magnet to lay the groundwork for precision measurements. 

Finally, we note that for applications of a ferromagnetic needle to measurement of non-magnetic torques, the shielding of external magnetic fields by the Meissner effect is an advantage. Such applications include searches for exotic spin-dependent interactions and dark matter \cite{safronova2018search,demille2017probing}. Furthermore, magnetometers with good spatial resolution are needed to study the superconductors themselves \cite{acosta2019color}.

\section{Experimental setup}
\begin{figure}
\centering
\includegraphics[width=8 cm]{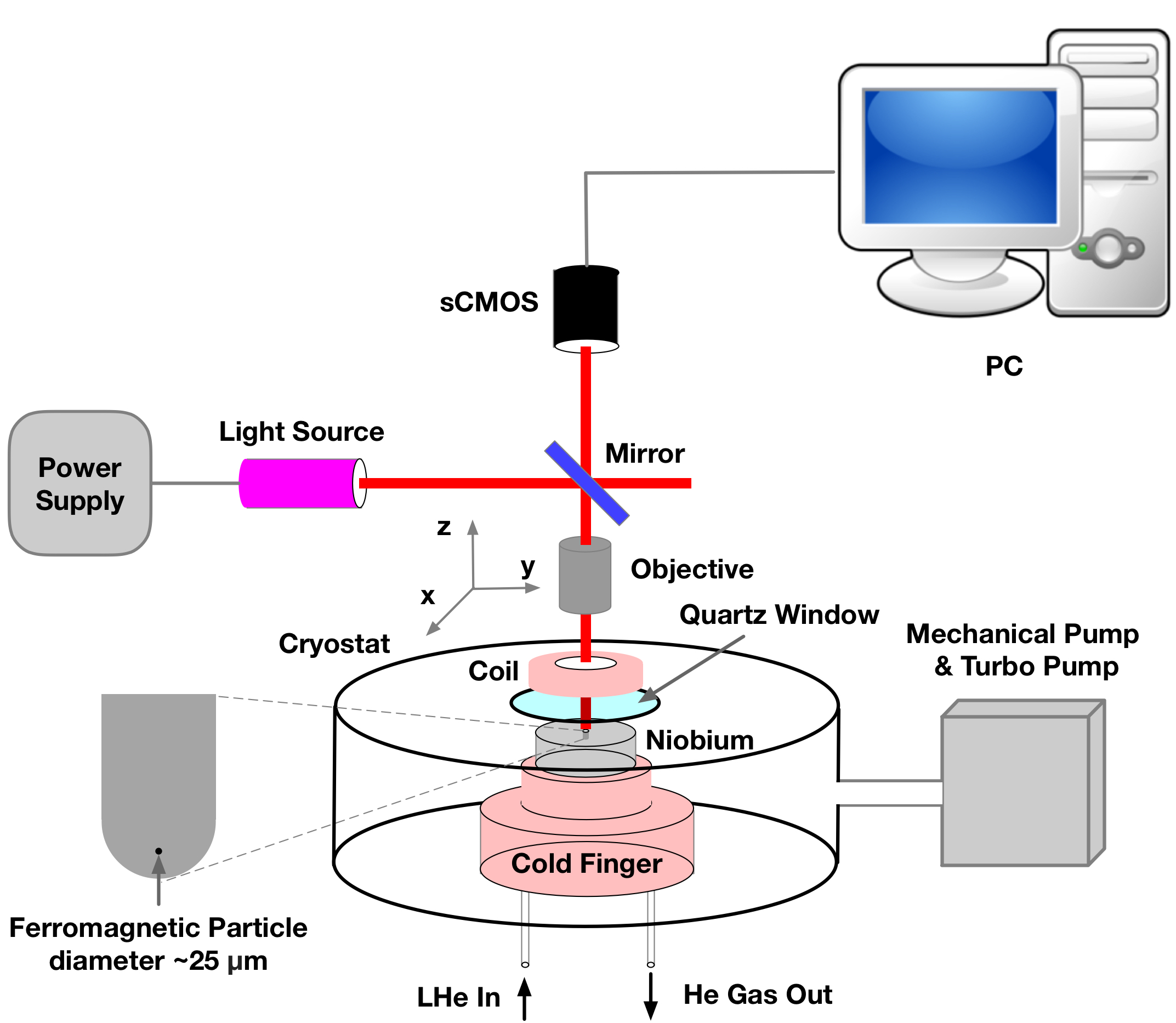}
\caption{\label{fig_setup} Experimental setup, the microscopy cryostat is pumped by the mechanical pump and turbo pump. The motion of the particle is recorded with the sCMOS camera above the cryostat. The coil is outside the cryostat. A detail view of the niobium well is shown at the left side.}
\end{figure}

The experimental setup is shown in Fig. \ref{fig_setup}. We use PrFeB ferromagnetic particles, whose shape is roughly spherical with a diameter of approximately 25 $\mathrm{\mu m}$. The particles are multi-domain and non-magnetized, so one particle can be easily separated from the other particles, then the chosen particle is magnetized by a strong permanent magnet and transferred into a niobium well (800 $\mu$m depth and 500 $\mu$m diameter) with a West Bond bonding machine integrated with a microscope and a micro-manipulator. The niobium well is made by electrical-discharge machining. The superconductor is placed into a microscopy cryostat. The cryostat chamber is pumped to a pressure of $10^{-5}$ Pa with a scroll pump and a turbo pump. The motion of the trapped levitating particle is characterized by long-lived oscillation modes of certain frequencies. The pump, while connected, excites near-resonant modes that can persist even when the pump is disconnected. When the cryostat reached the minimum temperature, the pumps were stopped and disconnected to reduce vibrations. White light from a Thorlabs 21AC halogen illuminator passes through a beam-splitting mirror, half of the light is reflected to pass through a 40$\times$ long-working-distance objective, and illuminates the magnetic particle and the superconductor. An Andor sCMOS (scientific Complementary Metal-Oxide-Semiconductor) camera is used for video recording with a frame rate faster than 250 frames per second. The superconductor is cooled down by liquid helium to 5 K, and the particle levitates when the niobium goes through the superconducting phase transition. We use a coil with approximately 1000 turns to provide an external magnetic field, whose radius and height are approximately 1.7 cm and 1.5 cm, respectively. The distance of the center of the coil from the niobium well is approximately 1.3 cm.

\begin{figure}
\centering
\includegraphics[width=8cm]{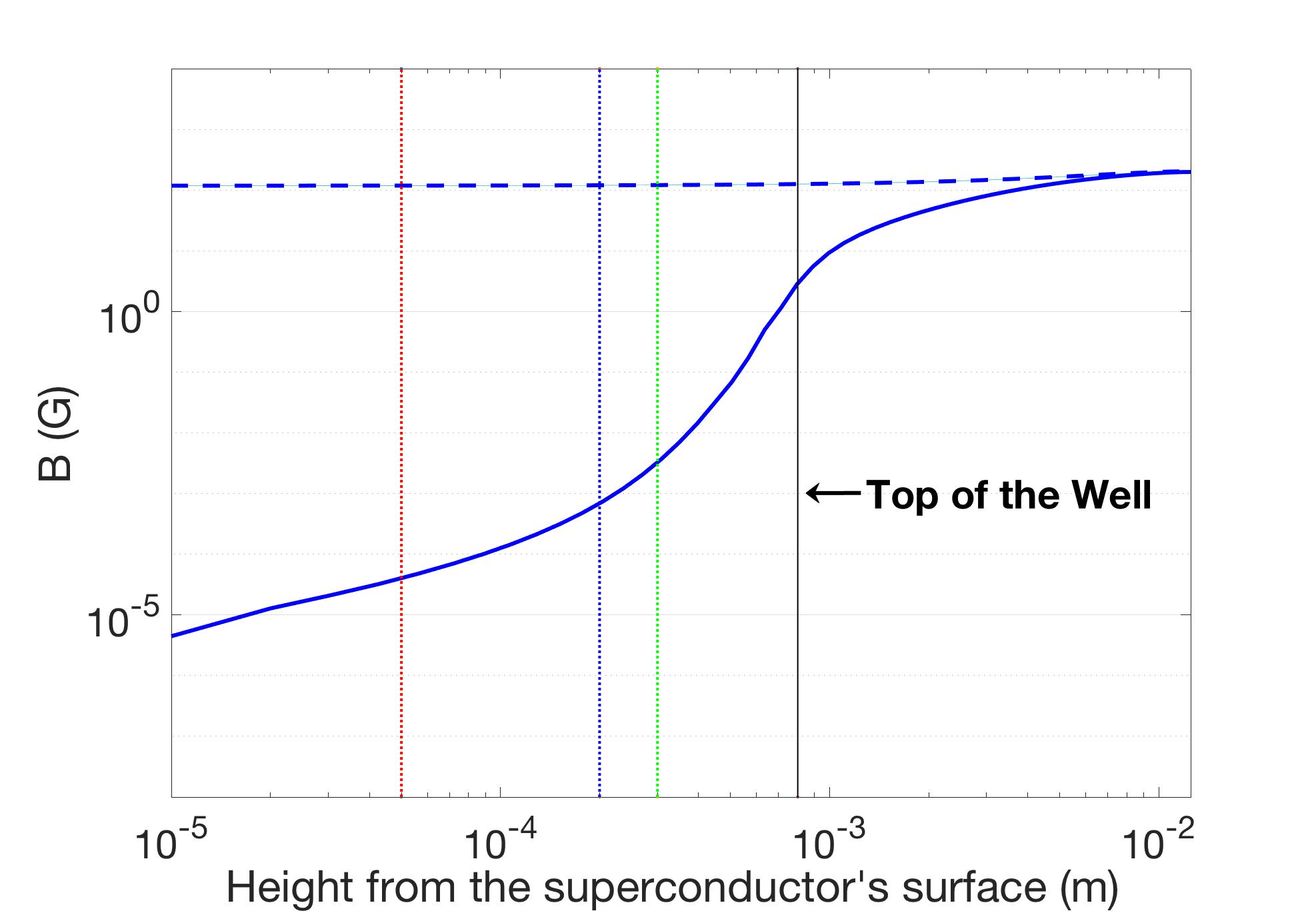}
	\caption{\label{fig_scsf} COMSOL simulation result for the shielding of the magnetic field from the coil near the surface of the superconductor well where the PrFeB particle is located. The applied magnetic field from the coil is approximately 100 G. The niobium was set to be ideally diamagnetic to simulate the Meissner effect \cite{finnemore1966superconducting}. The dashed and solid blue lines are the magnetic field distributions along the vertical symmetric axis when the superconductor is in the normal or superconducting states, respectively. The height is the distance of the particle from the bottom of the well. The shielding factor at 50 $\mu$m can be estimated based on the red dotted line, which is on the order of $10^6$. The shielding factors at 200 $\mu$m and 300 $\mu$m can be estimated in the same way, which are approximately $10^5$ and $10^4$, respectively. These heights are indicated by the blue and green dotted lines, respectively.}
\end{figure}

In the absence of any external applied field, the particle's equilibrium orientation and horizontal position is determined by the interaction between the particle and flux pinning sites (frozen image  \cite{kordyuk1998magnetic}) caused by impurities and structural defects in the superconducting well (both of which we are trying to keep at as low concentration as possible). At the same time, the particle’s equilibrium vertical position is determined by the much stronger interaction between the particle and its diamagnetic and frozen images as well as gravity. Consequently, the horizontal magnetic stiffness (elastic spring constant) is much smaller than the vertical magnetic stiffness. A vertically aligned external field from 0 to 100 G is applied to the sample, adjusted every two seconds in steps of about 4.5 G. This field is greatly attenuated by the superconductor at the position of the particle, as estimated by a finite-element COMSOL analysis detailed in Fig. \ref{fig_scsf}. The external coil and its diamagnetic image form a field configuration at the particle resembling that of an anti-Helmholtz coil, effectively cancelling the vertical field component but doubling any horizontal field component that exists due to misalignment or field gradients. When combined with much smaller horizontal magnetic stiffness, the primary effect of the external applied field is a shift in the equilibrium orientation and horizontal position of the particle, rather than a shift in its vertical equilibrium position. A step-jump in the applied field results in a non-adiabatic shift in the equilibrium position and orientation of the particle, exciting oscillatory linear motion and rotation in the particle, typically at a frequency of tens to on the order of one hundred of Hz. In some cases, the oscillatory motion may persist much longer than the two-second step interval. This motion is captured by the camera and the dynamics are analyzed.

\section{Results}
\subsection{Analysis of the video of the particle motion}

For each frame, a sequence of processing steps is performed in order to determine the position of the particle. First, the raw data are smoothed with Gaussian blurring to reduce the effect of noise on subsequent processing. The image is then segmented by thresholding at a value chosen such that boundaries of the image segments align with the regions having maximum gradient in brightness. Finally, the coordinates of the position of the particle are calculated by finding the centroid of the particle segment. This method produces position data that are largely insensitive to the flickering of the illumination.

Plotting the \textit{x} and \textit{y} position coordinates (defined by the camera) of the particle reveals that the coordinates oscillate with one or more frequencies along various directions during time intervals over which the magnetic field is held constant. These coordinates jump to a different average value when the magnetic field changes between steps, and oscillate with new frequencies and direction. These transition times are extracted by finding the peaks of the signal obtained from passing the position data through a digital low-frequency (1.0 Hz to 1.5 Hz) band-pass filter.

\begin{figure}[!ht]
\centering
	\includegraphics[width=8.5cm]{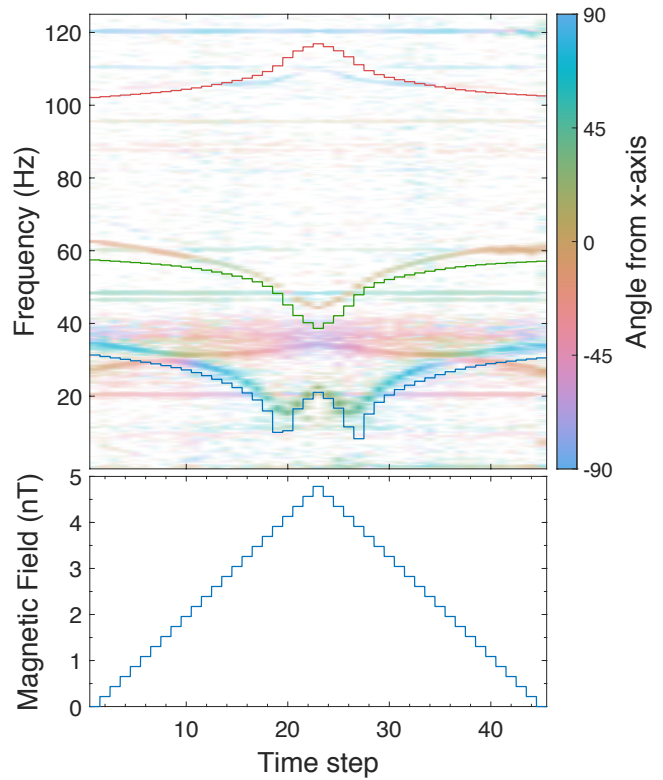}
	\caption{Measurements of the spectrum of the motion of a particle whose diameter is 25 $\mu$m and whose levitation height is approximately 80 $\mu$m. The objective is mounted on a Thorlabs 3-axis stage with differential micrometers for manual adjustments, the drive resolution is 0.5 $\mu$m. The objective has a high axial resolution. The objective focuses the image on the particle on the surface before it is levitated. When the particle is levitated, we refocus on the particle by precisely moving the stage vertically. The levitation height can be roughly measured in this way. In the upper figure, the color represents angle of the particle's vibration. The intensity of the lines is proportional to the spectral density of the signal. The constant frequency lines at 20, 40, 47, 95 Hz and 110 Hz were determined to be due to environmental vibrations/oscillations and not related to the particle motion (confirmed by observing the system with no ferromagnetic particle present). Near the 8th and 36th time step, the modes starting from 27 Hz and 35 Hz exhibit an avoided crossing. The bottom figure is the externally applied magnetic field felt by the particle taking into account the shielding due to the Meissner effect. Simulations of the motion of the particle above a superconductor, using the model in Fig. \ref{fig:frozen_image}, are overlain on the data plot. The vibrational modes are strong-coupled. The new eigenfrequencies [x (solid blue), y (solid green) and z (solid red)] are plotted by decoupling the original modes. The modes near 35 Hz, 62 Hz and 104 Hz are reproduced by the simulation. The free parameters in the simulation are $m_0$, $m_f$ and $\theta$.}
	\label{fig_result_b}
\end{figure}

\begin{figure}
\centering
\includegraphics[width=8.5cm]{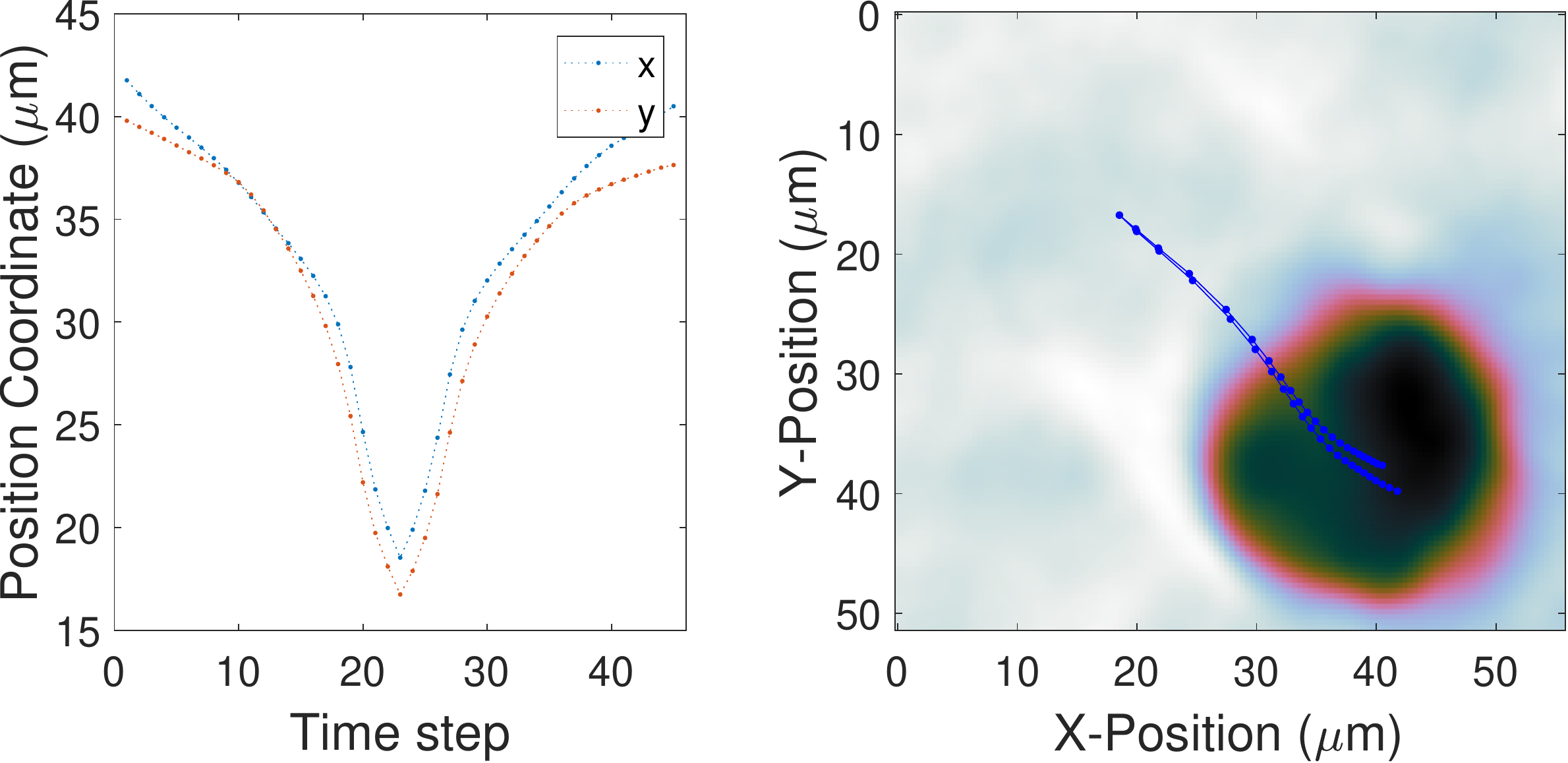}
\caption{\label{fig_result_c} Measurements of the position of the centroid of the particle (Left). The trajectory of the particle (Right), the black sphere is the levitated ferromagnetic particle, whose diameter is roughly 25 $\mu$m.}
\end{figure}

As a final step, the position data are spliced at the transition times into intervals of constant external magnetic field, and a spectral analysis is performed over each. These results are combined to form an image (Fig. \ref{fig_result_b}), similar to a spectrogram, in which more prominent frequencies are shown more brightly, with a hue that corresponds to the angle of the oscillations. The average coordinates for each timestep were also plotted against time as well as against each other (Fig. \ref{fig_result_c}).

%\begin{figure}
%	\centering
%	\includegraphics[width=10cm]{accel.jpg}
%	\caption{}
%	\label{fig:accel}
%\end{figure}

%\subsection{Avoided crossing--- Classical Landau-Zener transitions}
\subsection{Comparison between data and empirical model}
\begin{figure}
    \centering
    \includegraphics[width=8cm]{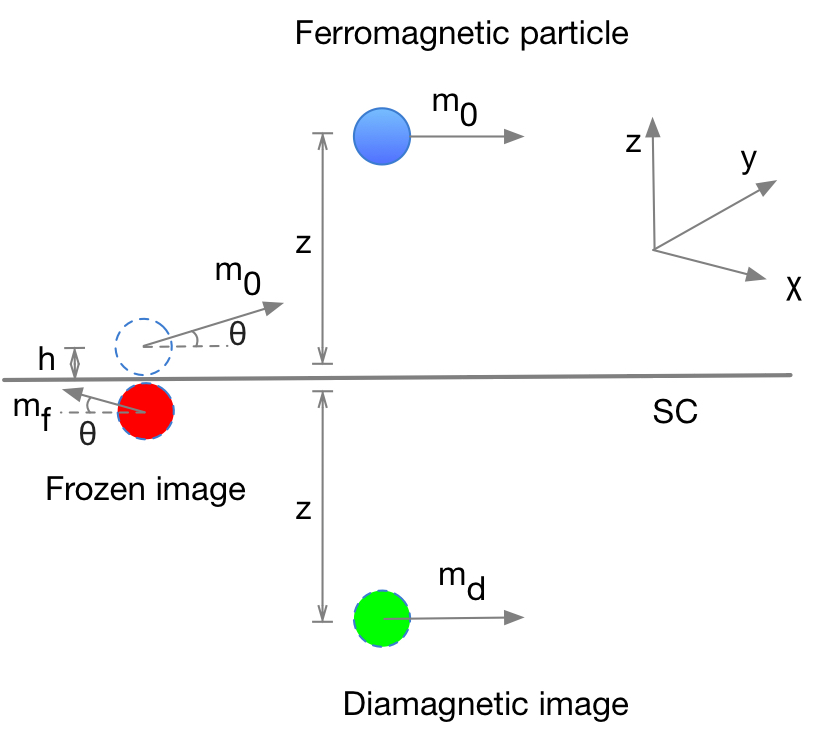}
    \caption{A model of the levitation mechanism of the ferromagnetic particle above a superconductor. The blue circle is the ferromagnetic particle, whose magnetization is $m_0$. The dashed blue circle is the initial position of the ferromagnetic particle during the cooling stage, and the frozen image $m_f$ (red dashed circle) was produced with a height $h$ inside the superconductor. The dashed green circle is the diamagnetic image, which is the mirror image of the particle.}
    \label{fig:frozen_image}
\end{figure}

The result of the video analysis is plotted in Fig. \ref{fig_result_b}. At the beginning of each time step, the magnetic field strength is changed and the step in the magnetic field excites the particle's motion. A dynamic model based on the frozen-image concept was developed to interpret the observations \cite{kordyuk1998magnetic}. As shown in Fig. \ref{fig:frozen_image}, there is a frozen image whose location depends on the cooling height $h$; and the location is fixed with respect to the particle's motion. Due to the Meissner effect, there is also a diamagnetic image \cite{giaro1990correct}, which is a mirror image of the particle; the diamagnetic image follows the motion of the particle. When the particle moves horizontally away from the originally stable position, the frozen image provides an attractive force to ``drag" the particle back to the original position, and the diamagnetic image provides a repulsive force to keep the magnetic particle away from the surface of the superconductor. The simultaneous presence of a frozen image and a diamagnetic image, and the displacement from the original particle position make the magnetic stiffness \cite{qin2002calculation} between x, y and z correlated, hence the system represents (at a minimum) a three-mode coupled problem. In a strongly coupled system, avoided crossings can occur \cite{novotny2010strong}, and indeed an avoided crossing is observed at the $\mathrm{8^{th}}$ and $\mathrm{36^{th}}$ time steps shown in Fig. \ref{fig_result_b}. The dynamic model that we use is described in the supplementary material \cite{supplemental}.

There are a number of magnetic-field-independent oscillatory signals observed in the data presented in Fig. \ref{fig_result_b}. Auxiliary measurements without the magnetic particle present (see supplemental material \cite{supplemental}) have shown that these signals are unrelated to the motion of the magnetic particle. The 120 Hz signals in Fig. \ref{fig_result_b} are the harmonics of the line frequency coupled through the light source. The 60 Hz signals are the line frequency coupled through the coils' current supply. As discussed in the supplemental material \cite{supplemental}, the constant signals at frequencies of 20, 40, 47, 95 and 110 Hz signals are the result of background noises that originate from the vibrational modes of the apparatus. All the spectral lines in Fig. \ref{fig_result_b} whose frequencies change as the applied magnetic field is changed are identified with the motion of the particle. Their frequencies at the zero time step are 27, 35, 62 and 104 Hz (we will identify the modes by their frequencies at zero time step, which are the modes of the particle with no applied magnetic field). These modes represent three translational vibrations and a rotational mode (the particle is not ideally spherical, the centroid moves when the particle rotates). The simulation results for the vibrational modes as a function of the magnetic field are shown in Fig. \ref{fig_result_b}, and are based on the equations discussed in supplementary material \cite{supplemental}. In the simulation, because the superconductor was cooled down with the ferromagnetic particle on the surface, the cooling height is assumed to be equal to the particle's radius ($h= 12.5$ $\mu$m), the remanence of the particle is 3000 G, and the remanence of the frozen image is 20 G. This indicates that the pinning field due to trapped flux originating from impurities and structural defects is relatively small compared with that expected in hard superconductors. In order to clearly observe the precession, for a spherical ferromagnetic particle with radius $r$, the magnetic field must be smaller than $N\hbar^2/(g\mu_B I)=5\hbar^2/(2g\mu_Bm_ar^2)$, where $\mu_B$ is the Bohr magneton, $m_a$ is the atomic mass in kg, $g \approx 1$ for cobalt. For a ferromagnetic particle with a diameter of 25 $\mu$m, the threshold magnetic field is approximately 2 $\mu$G. The magnetic field from the frozen image felt by the particle is estimated to be $\frac{2}{3}\frac{B_r}{(2+h_0/r)^3} \approx$ 50 mG. Thus, the remanence of the frozen image is strong enough to prevent us from clearly observing the particle's precession. 

Comparing the experimental result to the simulation in Fig. \ref{fig_result_b}, many of the qualitative features of the data are reproduced. However, the 27 Hz mode, which exhibits an avoided-crossing with the 35 Hz mode is missing in the simulation result. This mode could be a rotational mode instead of a translational mode \cite{cansiz2005translational}. This simulation only includes the translational modes.

\begin{figure}
\centering
	\includegraphics[width=8 cm]{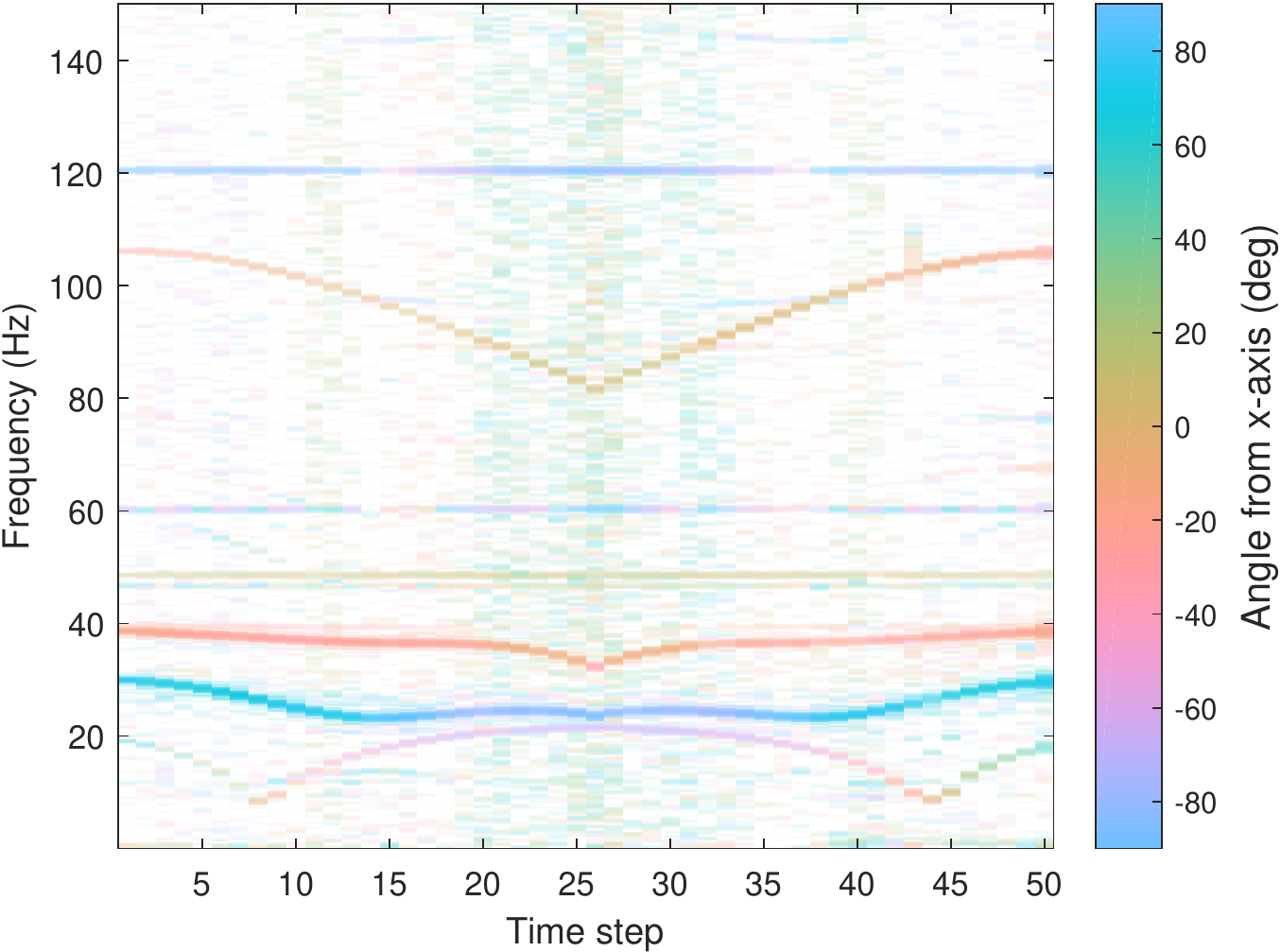}
	\caption{Measurements of the spectrum of motion of a ferromagnetic particle whose diameter is approximately 30 $\mu$m. The levitation height of the particle is approximately 200 $\mu$m. Again, the 60 Hz and 120 Hz signals in the spectrum are the line frequency and its harmonic.}
	\label{fig_result_200}
\end{figure}

\begin{figure}
\centering
	\includegraphics[width=8cm]{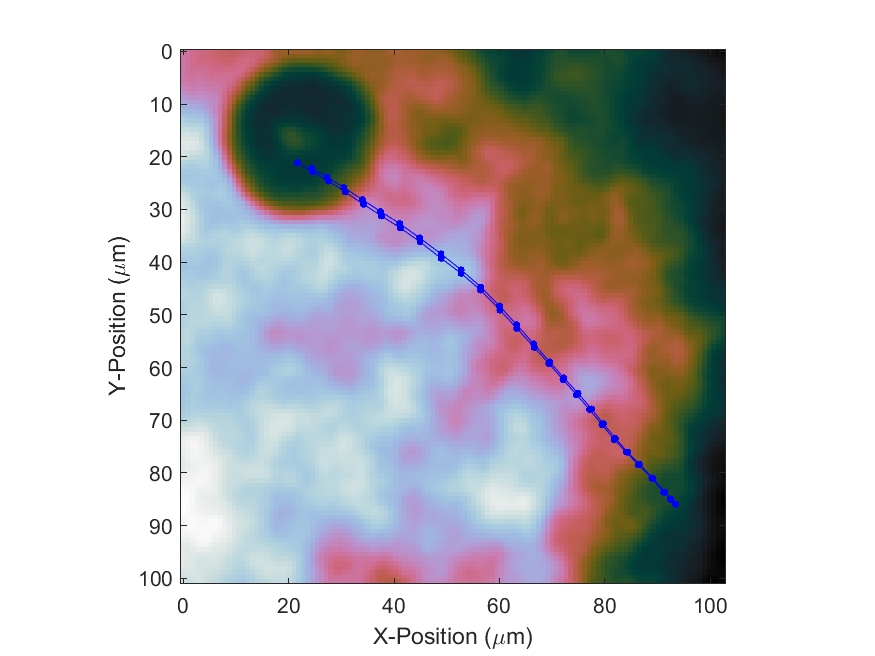}
	\caption{Trajectory of the 30 $\mu$m particle. Note the larger path length relative to the 25 $\mu$m particle shown in Fig. \ref{fig_result_c} (Right). The greater levitation height (200 $\mu$m vs 50 $\mu$m) results in reduced screening of the external applied field, and thus a greater offset of the horizontal equilibrium position.}
	\label{fig:i628_trajectory}
\end{figure}

The niobium surface was cleaned with acetone, thereby reducing the impurity level; consequently the flux pinning was reduced. The experiment was redone with a 30 $\mu$m-diameter particle; the result is plotted in Fig. \ref{fig_result_200}, and the trajectory is plotted in Fig. \ref{fig:i628_trajectory}. A levitation of approximately 200 $\mu $m was achieved. The modes of the particle's vibration are 18 Hz, 29 Hz, 38 Hz and 106 Hz at the zero time step. Comparing Fig. \ref{fig_result_b} with Fig. \ref{fig_result_200}, the lowest-frequency mode (27 Hz mode for Fig. \ref{fig_result_b}, 18 Hz mode for Fig. \ref{fig_result_200}) has similar trend, the frequency first decreases then increases with an amplitude sweeping DC magnetic field, whose amplitude is linearly increased.

\subsection{Q-factor and dampings}
There are several damping sources for a levitated particle, including molecular collision damping, radiative damping, and eddy current damping. In our setup the power of the light used to observe the particle is sufficiently weak so that radiative damping cannot be the dominant damping term. Therefore, damping from the molecular collisions and eddy currents are most likely the dominant effects, and are discussed here.

The drag force on a spherical ferromagnetic particle caused by collisions with gas molecules can be written as \cite{de2013drag}
\begin{equation}
	F_d=\frac{4}{3}\rho_g \pi r^2 v_{p} \sqrt{v_{g}^2+v_{p}^2},
\end{equation}
where $\rho_g=n m_g$ is the mass density of the gas, $n$ is the number density of the gas, where $n=P_{g}/(k_B T)$, $P_{g}$ is the gas pressure, $m_g$ is the mass of the molecule, $T$ is the temperature, $v_p$ is the velocity of the particle, $v_g$ is the average thermal velocity of the gas molecules, where $v_g=\sqrt{3k_BT/m_{g}}$. In our experiment, $v_p\ll v_g$. We estimate the average velocity of the particle $\bar{v_p} \approx \frac{1}{2}\omega A_m$ and the energy dissipated per cycle, obtained by integrating the drag force from Eq. (1)  over the particle trajectory, is 
\begin{equation}
	E_g \approx \bar{F_d}\times 4A_m=\frac{8}{3}\rho_g\pi r^2v_g\omega A_m^2,
\end{equation}
where $A_m$ is the maximum amplitude of the particle's vibration.

%The angular momentum imparted from the collisions with the residual gas molecules is 
%\begin{equation}
%	L_{c}\approx \frac{2m_{He}v_{He}r}{5},
%	\label{}
%\end{equation}
%where $m_{He}$ is the mass of He, $v_{He}$ is the average thermal velocity of He, which equals $\sqrt{3k_BT/m_{He}}$, $r$ is the radius of the ferromagnetic particle, the factor of 2/5 arises from averaging over the angle and location of impact. The collision rate is 
%\begin{equation}
%	\Gamma_{c} \approx \frac{nAv_{He}}{4},
%\end{equation}
%where $n$ is the density of He gas, which can be calculated from $(P_{He} N_{A})/(RT)$, $P_{He}$ is the gas pressure of He, $R$ is ideal gas constant, $N_{A}$ is Avogadro constant and $A= \pi r^2$.
%The energy dissipated per cycle is 
%\begin{equation}
%	E_{c}=\frac{\frac{L_{c}^2}{2m}\Gamma_{c}}{f}.
%\end{equation}
The total energy stored is 
\begin{equation}
	E_{v}=\frac{1}{2}m\omega^2A_m^2.
\end{equation}
Thus, the q-factor equals
\begin{equation}
	Q_{g}=2\pi \frac{E_{v}}{E_{g}}=\frac{\pi \omega}{6}\frac{\rho r}{P_g}\sqrt{\frac{3k_BT}{m_g}}=\frac{\pi \omega}{6}\frac{\rho v_g r}{P_g}.
\end{equation}
The density of the particle $\rho \approx 7\times 10^3 ~\mathrm{kg/m^3}$, we assume the background gas is He, $\omega$ is assumed to be $2 \pi \times $ 40 Hz, the temperature $T \approx 6$ K, and so the Q-factor determined by collisions with background gas molecules equals $10^{8}$, which is not a limitation under our experimental conditions.

In order to estimate the eddy current damping, the particle is treated as a current loop. When the particle is precessing in the plane horizontal to the surface of the superconductor, there will be no eddy current damping from the diamagnetic image, because the diamagnetic image always rotates with the particle, and so there is no change of the magnetic flux from the diamagnetic image. However, the oscillation along the vertical direction will cause eddy current damping. (Note that this mechanism may be useful for realizing a precessing needle magnetometer in the future, since all the rotational and vibrational motion outside of the plane parallel to the superconductor surface will damp out, leaving only the precession motion in the plane parallel to the surface of the superconductor, which is of principal interest). Here the damping of rotational motion in a plane perpendicular to the superconductor surface is studied. If the current loop flips with frequency $\omega$, the changing magnetic-flux-induced voltage can be written as
\begin{equation}
    U=\frac{d\Phi}{dt}=\frac{d[BA sin(\theta(t))]}{dt}=BAcos(\theta(t))\omega.
\end{equation}
The resistance of the spherical particle can be written as 
\begin{equation}
    R=\frac{\pi r}{1/2\pi r^2\sigma}=\frac{2}{\sigma r}.
\end{equation}
The energy dissipated per cycle is 
\begin{equation}
\begin{aligned}
    E_{e}&=\int_{0}^{T}\frac{U^2}{R}dt=\int_{0}^{T}\frac{B^2A^2\omega^2}{R}\frac{1+cos(2\theta(t))}{2}dt\\
    &=\frac{\pi^3 B^2 r^4\omega}{R},
\end{aligned}
\end{equation}
and the total energy stored is 
\begin{equation}
    E_{r}=\frac{1}{2}I\omega^2.
\end{equation}
Thus, the Q-factor equals
\begin{equation}
    Q_c=2 \pi \frac{E_{r}}{E_{e}}=2\pi \times \frac{\frac{1}{2}I\omega^2}{\frac{\pi^3 B^2 r^4\omega}{R}}=\frac{16 \rho \omega}{15 \pi \sigma B^2}\approx\frac{\rho\omega}{\pi\sigma B^2}.
\end{equation}
If we assume the electrical conductivity\cite{stipe2001electron} $\sigma=10^7 ~\mathrm{\Omega^{-1}\cdot m^{-1}}$, the magnetic field felt by the particle from the diamagnetic image and the frozen image $B=$ 4 G, $\omega \approx 2\pi \times 40$ Hz, then the Q-factor equals $3\times 10^5$. The particle was actually levitated from the surface of the superconductor, which was cooled down to 5 K. The electrical conductivity could be even larger in this case, and so the q-factor limited by the eddy current damping could be even smaller. This indicates that eddy currents may be a dominant source of damping in our setup.

\begin{figure}
\centering
	\includegraphics[width=8cm]{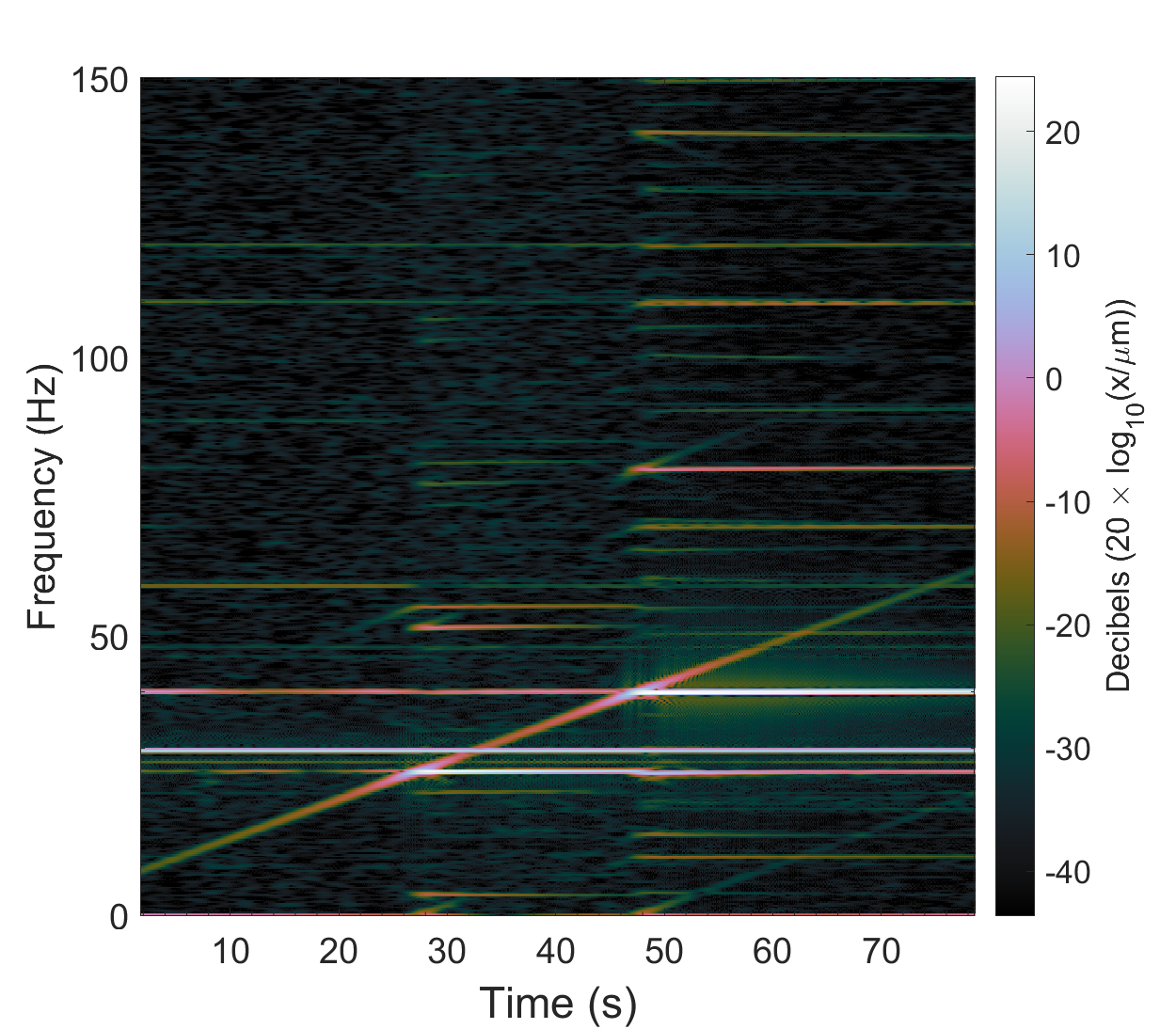}
	\caption{The frequency of the oscillating magnetic field was swept from 5 Hz to 60 Hz, the 26 Hz and 40 Hz modes and their harmonic modes were exited. The diagonal line is the response to the swept AC magnetic field. Here the pumps were not disconnected from the cryostat. The 29.5 Hz mode was excited by vibration caused by the scroll pump, whose rotation speed is 1770 rpm (= 29.5 Hz). The color bar on the right side indicates the amplitude of the motion.}
	\label{fig:0628g_centroid_g_spectrogram_raw}
\end{figure}

\begin{figure}
\centering
	\includegraphics[width=8cm]{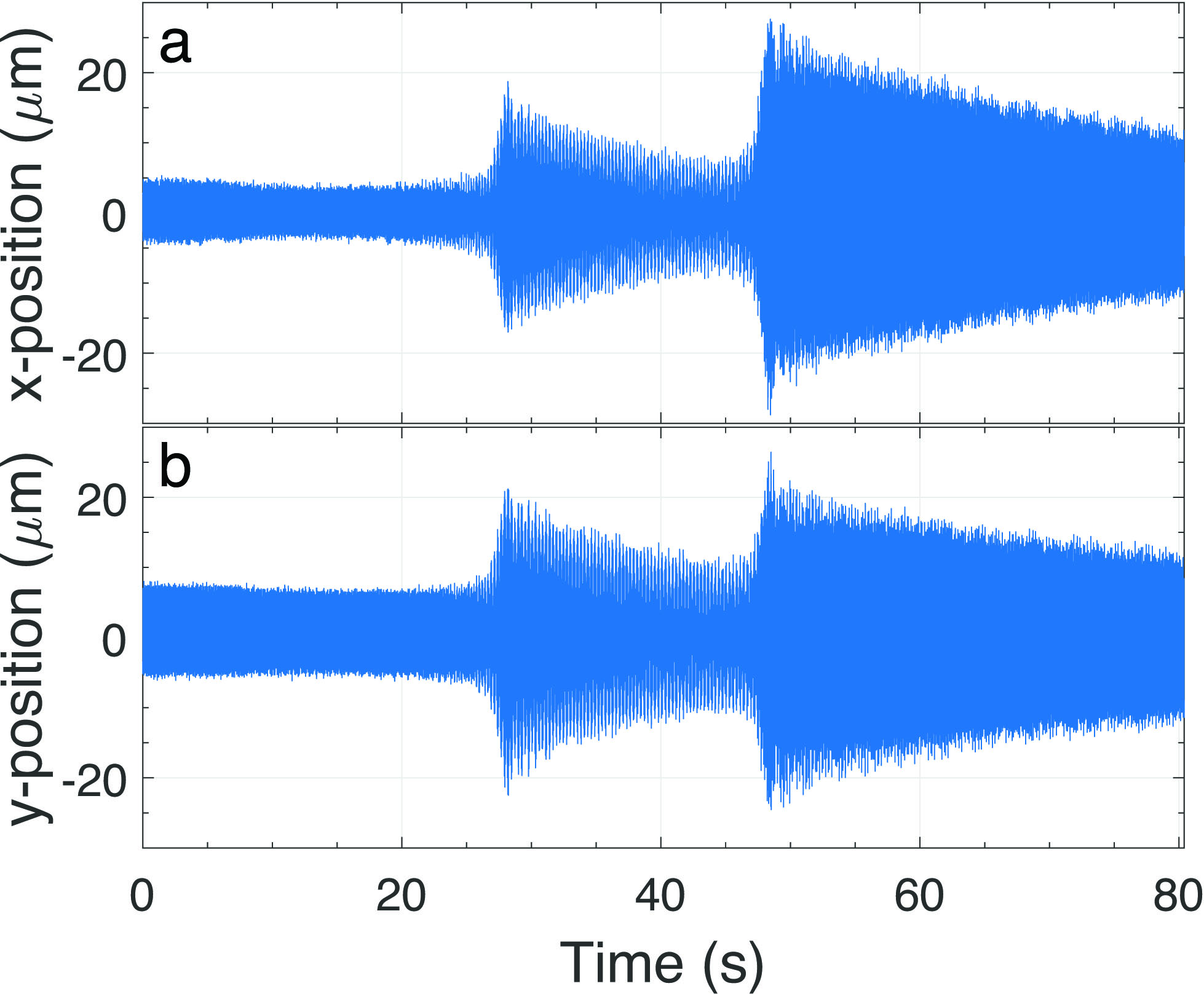}
	\caption{Measurements of the x-coordinate (a) and y-coordinate (b) of the particle's position as a function of time. The particle's motion was excited at the time of approximately 28 s and 47 s.}
	\label{fig:0628g_centroid_h_amplitude}
\end{figure}

\begin{figure}
	\centering
	\includegraphics[width=8cm]{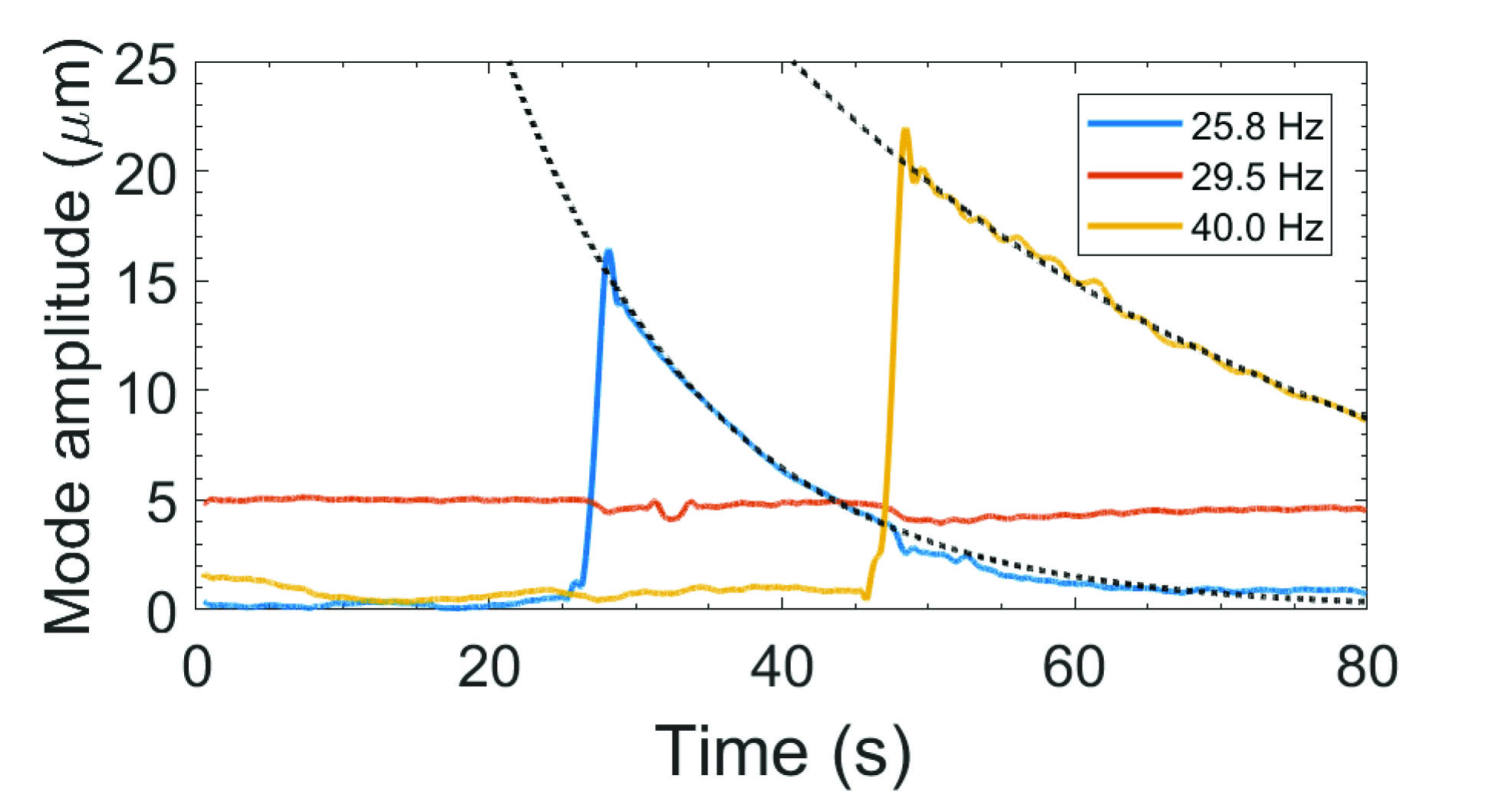}
	\caption{Measurements of the amplitude of the vibrations for each mode as a function of the sweeping time. The scroll pump was on, the 29.5 Hz signal from the scroll pump (1770 rpm) still can be seen in figure, which excited the particle motion. We calculated the decay time by fitting the amplitude curve exponentially. The Q-factor is approximately $1\times10^3$, $4\times10^3$ for 26 Hz and 40 Hz modes, respectively.}
	\label{fig:0628g_centroid_f_decibels_1D}
\end{figure}

\begin{table}
\caption{The measurement results of the Q-factor}
\begin{tabular}{cccc}
\hline
  & Sweeping Frequency Range   & Mode & Q-factor \\
\hline
(a)  & 5-60 Hz & 26 Hz & $1\times10^3$\\
	&	&40 Hz &$4\times10^3$ \\
\hline
(b) & 5-150 Hz & 38 Hz & $4\times10^3$\\
& &76 Hz & $7\times10^3$  \\
\hline
(c) & 10-60 Hz & 38 Hz & $6\times10^3$\\
\hline
(d) & 20-60 Hz & 38 Hz & -  \\
	&	&106 Hz & $2\times10^4$	\\
\hline
(e) & 35-45 Hz & 38 Hz & $5\times10^4$  \\
\hline
\label{table:q-factor}
\end{tabular}
\end{table}
In the experiment, the Q-factor was measured by continuously linearly sweeping the frequency of an oscillating magnetic field, then measuring the decay time. The experimental results are shown in Figs. \ref{fig:0628g_centroid_g_spectrogram_raw}, \ref{fig:0628g_centroid_h_amplitude} and \ref{fig:0628g_centroid_f_decibels_1D}. We noticed that the 30 Hz mode seems to have a very high Q-factor, the mode is close in frequency to one of the apparatus's vibrational frequencies, which means it could be continuously excited by the apparatus' vibration. The measurement results of the Q-factor are listed in the Table \ref{table:q-factor}. For the data set (a), the particle levitated from a different site of the superconductor surface, hence the eigenmodes are different. All the other datasets (b-e) were measured with the particle levitated from the same site, and the datasets are listed according to the time order. The Q-factor of the 38 Hz mode increases from (b) to (e), possibly due to progressive heating and increasing resistivity of the levitating particle due to the lamp, consistent with the primacy of eddy current damping. The Q-factor of 38 Hz mode is missing in (d), because the 38 Hz did not fully decay when there was a new excitation, the 106 Hz mode was strongly excited instead, these modes were coupled. The Q-factors from these experimental results ranged from $10^3$ to $5\times 10^4$.

%
%\subsection{Excitation of the particle's motion}
%In this experiment, there is no excitation magnetic field applied to excite the particle's modes. The particle's motion was excited by the vibrations from the mechanical pump and the turbo pump. The spectrum of the apparatus' vibration is shown in 

%\begin{figure}
%    \centering
%    \includegraphics[width=12cm]{628f_centroid_g_spectrogram_raw.png}
%    \caption{The frequency of the oscillating magnetic field was swept from 5 Hz to 150 Hz,}
%    \label{fig:628f_centroid_g_spectrogram}
%\end{figure}
%
%\begin{figure}
%    \centering
%    \includegraphics[width=12cm]{628f_centroid_j_resamplitude.png}
%    \caption{Caption}
%    \label{fig:628f_centroid_j_resamplitude}
%\end{figure}

\section{Approach to magnetometry beyond the standard quantum limit}
Due to the non-negligible magnetic flux pinning in the superconductor, the particle was not in the precession-dominant regime. However, flux pinning can be significantly reduced by zero-field cooling, which involves dropping the particle after the superconductor is cooled below the critical temperature. If the flux pinning can be sufficiently suppressed, the precession of a ferromagnetic particle could be clearly observed. Furthermore, because of the superconducting shielding effect, the levitated needle magnetometer, unlike SERF magnetometers and SQUID magnetometers \cite{dang2010ultrahigh, storm2017ultra}, will not be limited by Johnson thermal noise. The preccessing ferromagnetic needle magnetometer can only work in fields smaller than the threshold magnetic field. The diamagnetism of the superconductor can shield out most of the external magnetic field, which is useful, for example, in searching for dark matter  or exotic spin-dependent interactions \cite{safronova2018search,demille2017probing}. Furthermore, a superconducting flux transformer proposed in \cite{ichkitidze2012superconducting, wang2018application} can help the needle magnetometer to measure external oscillating magnetic fields. 

\section{Challenges Ahead}

The flux pinning induced frozen image produces a strong torque on the levitated particle, which exceeds the critical torque associated with the transition between precession and oscillation of the ferromagnetic particle. Future work will involve developing methods to suppress or counteract the flux pinning in order to reach the regime where precession of the ferromagnetic particle can be clearly observed and studied.

Even if flux pinning is fully suppressed, the particle may still carry out vertical vibrational and rotational motions, complicating  detection of the precession mode. Replacing the camera with a SQUID or NV-diamond magnetometer to detect the motion of the particle could enable more precise measurement of the ferromagnetic particle dynamics and enable observation and study of the precessional motion.

\section{Conclusion}
The dynamics of a ferromagnetic particle levitated above a superconductor were studied. A frozen image model was used to explain the dynamics of the particle's motion. The simulation results qualitatively reproduce the experimental results. This research lays the foundation for future research and development of a precessing ferromagnetic needle magnetometer.

\section{Acknowledgments}
The authors are sincerely grateful to  Prof. John Clarke and Sylvia Lewin (UC Berkeley Physics Department), Alexander Wilzewski (JGU Mainz), Dr. Jianmei Huang and Dr. Dylan Lu (UC Berkeley Chemistry Department), Prof. Ming Zeng (Beihang Univ.) for the help in the preliminary study. This work was supported by the Simons and Heising-Simons Foundations and a grant from the DFG through the DIP program (FO703/2-1). DFJK acknowledges the support of the U.S National Science Foundation under grant No. PHY-1707875.

\section*{References}
% \nocite{*}
\bibliography{reference}

\end{document}